\documentclass[11pt]{revtex4-1}

\usepackage{graphicx}
\usepackage{dcolumn}
\usepackage{color}
\usepackage{bm}
\usepackage{amsmath}
\usepackage{amsfonts}
\usepackage{amssymb}

\input epsf

\begin{document}

\title{A Cosmological Study in Massive Gravity theory}

\author{Supriya Pan\footnote{span@research.jdvu.ac.in}}
\affiliation{Department of Mathematics, Jadavpur University, Kolkata-700 032, India.}

\author{Subenoy Chakraborty\footnote{schakraborty@math.jdvu.ac.in}}

\affiliation{Department of Mathematics, Jadavpur University, Kolkata-700 032, India.}{}

\begin{abstract}
A detailed study of the various cosmological aspects in massive gravity theory has been presented in the present work. For the homogeneous and isotropic FLRW model, the deceleration parameter has been evaluated, and, it has been examined whether there is any transition from deceleration to acceleration in recent past, or not. With the proper choice of the free parameters, it has been shown that the massive gravity theory is equivalent to Einstein gravity with a modified Newtonian gravitational constant together with a negative cosmological constant. Also, in this context, it has been examined whether the emergent scenario is possible, or not, in massive gravity theory. Finally, we have done a cosmographic analysis in massive gravity theory.\\


Pacs No: 04.50.-h, 98.80.-k, 95.36.+x.

\end{abstract}
\maketitle
\section{Introduction}
In classical field theory, a basic question is whether it is possible to introduce a consistent extension of general relativity for massive graviton, or not. In fact, a small graviton mass may have significant physical interest, notably, for the cosmological constant problem. An attempt for a covariant theory was initiated long back in 1939 by Fierz and Pauli (FP) \cite{FP1}. They introduced a quadratic mass term $m^2 (h_{\mu \nu}h^{\mu \nu}-h^2)$ for linear gravitational perturbation $h_{\mu \nu}$ to the action, and, as a result, there is a violation of gauge invariance in general relativity. Also, the linear theory with FP-mass does not recover General Relativity in the massless limit $m\rightarrow 0$, a contradiction with solar system test due to the van Dam--Veltman--Zakharov (VDVZ) discontinuity \cite{VDV1, Zakharov1}. This discontinuity can be overcome by introducing nonlinear interactions with the help of Vainshtein mechanism \cite{Vainshtein1}. But elimination of VDVZ potential led inevitably to the existence of Boulware--Deser (BD)-ghost \cite{Boulware1}, and, as a result, the theory becomes unstable \cite{Arkani-Hamed1, Creminelli1, Deffayet1, Gabadadze1}.\\
Then after a long gap, very recently, a two parameter family of nonlinear generalization of the FP theory was formulated by de Rham, Gabadadze and Tolley (dRGT) \cite{deRham1, deRham2}. This model of massive gravity is in the context of extra dimensions, and it modifies general relativity at the cosmological scale. Here, the BD ghosts are eliminated in the decoupling limit to all orders in perturbation by constructing a covariant nonlinear action \cite{Hassan1, Hassan2, deRham3, deRham4} (see ref. \cite{Hinterbichler1} for a review), and is free from tames issues which have plagued earlier attempts at a given mass to the graviton. In the context of studying the nature of perturbations on the given backgrounds, Fasiello and Tolley \cite{FT2012} have investigated the Higuchi bound for classical stability of dRGT theory with the FLRW reference metric, and, in contrast to earlier investigations \cite{GS2010, BDH2010, BDH2011, BCNP2009}, it is found that the bound is independent of the precise form of matter. Also, it has been shown \cite{FT2012} that the free parameters do not change the qualitative picture of the Higuchi-Vainshtein analysis except for a specific cosmological epoch. Furthermore, it is found that the Higuchi-Vainshtein tension can not be relaxed in the parameters space for stability of the dRGT theory if both the reference and the dynamical metrics are FLRW, i.e., the cosmological solutions (in massive gravity) with a general FLRW reference metric are not phenomenologically viable due to the tension with an operative Vainshtein mechanism \cite{FT2012}.\\
At present, there are wide classes of applications of dRGT theory, particularly, in cosmology \cite{DAmico1, Gumrukcuoglu1, Gumrukcuoglu2, DeFelice1, Gumrukcuoglu3, Comelli1, Comelli2, Cardone1, Gratia1, DAmico2, Saridakis1}, in Black holes, and, in spherically symmetric solutions \cite{Koyama2, Koyama3, Nieuwenhuizen1, Gruzinov1, Comelli3, Berezhiani1, Brihaye1, Cai2} and its relation with bimetric gravity \cite{Damour1, Hassan3, Hassan4, vonStrauss1, Hassan5, Volkov1, Volkov2, Baccetti1, Baccetti2}. In the present work, various cosmological aspects have been studied in the perspective of the massive gravity theory. The paper is organized as follows: Section II deals with basic equations in massive gravity theory. Deceleration parameter has been evaluated, and possible transitions have been studied in section III. Section IV discusses on the equivalence of massive gravity and Einstein gravity. In Section V, we proposed an emergent scenario. Section VI is devoted for a detail cosmographical analysis. Finally, in Section VII, we have briefly discussed our results.

\section{Basic equations for massive gravity}

Massive gravity theory has an effective field theoretic prescription given by Einstein gravity together with the covariant FP mass term. In dRGT model, the action is a functional of the metric $g_{\mu \nu}(x)$ and four spurious scalar fields $\phi^a (x), a= 0, 1, 2, 3$, which are introduced to give a manifestly diffeomorphism invariant description \cite{Arkani-Hamed1}. One defines a covariant tensor $H_{\mu \nu}$ as:

\begin{equation}
g_{\mu \nu}= \partial_\mu \phi^a \partial_\nu \phi^b \eta_{ab}+ H_{\mu \nu},\label{eqn1}
\end{equation}

where $\eta_{ab}= diag(-1, +1, +1, +1)$, the usual Minkowskian metric is defined in the co-ordinate system defined by $\phi^a$'s. So, gravity in this formulation is described by the tensor $H_{\mu \nu}$ propagating on Minkowski's space. In the unitary gauge, all the four scalars $\phi^a (x)$ are frozen, and, equal to the corresponding space-time co-ordinates $\phi^a (x)= x^\mu \delta _{\mu} ^a$. However, sometimes, it may be helpful to use a non-unitary gauge in which $\phi^ a (x)$'s are allowed to fluctuate. So, a covariant Lagrangian density for massive gravity can be written as \cite{deRham1, deRham2}

\begin{equation}
L= \frac{1}{16 \pi G} \sqrt{-g} \left[R+ m_g ^2 U(g, H)\right],\label{eqn2}
\end{equation}

where $R$ is the usual Ricci scalar, $m_g$ is the mass of graviton and $U$, the nonlinear higher derivative term for massive gravity is a potential for the graviton which modifies the gravitational sector. The explicit form for $U$ is given by

\begin{equation}
U= U_2+ \alpha_3 U_3+ \alpha_4 U_4,\label{eqn3}
\end{equation}

where, $\alpha_3$, $\alpha_4$ are dimensionless parameters, and, $U_i$'s ($i= 2, 3, 4$) have the expressions
\begin{eqnarray}
U_2&=& [\kappa]^2- [\kappa^2],\label{eqn4}\\
U_3&=& [\kappa]^3-3[\kappa][\kappa^2]+2[\kappa]^2,\label{eqn5}\\
U_4&=& [\kappa]^4-6[\kappa^2][\kappa]^2+8[\kappa^3][\kappa]-6[\kappa^4].\label{eqn6}
\end{eqnarray}

Here, $\kappa ^\mu _\nu$ is defined by the four Stueckelberg fields $\phi^a$ (to restore general covariance \cite{Arkani-Hamed1}) as

\begin{equation}
\kappa ^\mu _\nu= \delta_\nu ^\mu- \sqrt{g^{\mu \sigma} f_{ab} \partial_\sigma \phi ^a \partial_\nu \phi^b},\label{eqn7}
\end{equation}

and the square bracket stands for the trace of the corresponding quantity within it, i.e.,

\begin{equation}
[\kappa]= \kappa^\mu _\mu,~~~~~[\kappa^2]= \kappa_\rho ^ \mu \kappa_\mu ^ \rho,\label{eqn8}
\end{equation}

and, so on.. Also, we write

\begin{equation}
\Sigma_{\mu \nu}= \partial_\mu \phi^a \partial_\nu \phi^b f_{ab},\label{eqn9}
\end{equation}

as the symmetric (0, 2) tensor defined by the Stueckelberg fields $\phi^a$. Normally, the reference metric $f_{ab}$ is chosen as the usual Minkowski metric $\eta_{ab}$. Now, for convenience, if we choose the unitary gauge $\phi^a (x)= x^\mu \delta _{\mu} ^a$, then the metric tensor $g_{\mu \nu}$ stands for the observable describing the 5 degrees of freedom of the massive graviton.

\section{Massive gravity in FLRW model}

The line element for Friedmann--Lema\^{\i}tre--Robertson--Walker (FLRW) geometry of arbitrary spatial curvature is given by

\begin{eqnarray}
ds^2&=& g_{\alpha \beta} dx^ \alpha dx^\beta= -N^2 (t) dt^2+ a^2 (t) \gamma_{ij} dx^i dx^j,\label{eqn10}
\end{eqnarray}

with spatial metric $\gamma_{ij}$ in terms of spherical co-ordinates as

\begin{equation}
\gamma_{ij} dx^i dx^j= \frac{dr^2}{1- K r^2}+ r^2 (d \theta^2+ \sin^2 \theta d \phi^2), \label{eqn11}
\end{equation}

where $K= 0, +1$, or $-1$ refers to flat, closed, or, open models of the universe respectively.\\

In the present work, we choose the reference metric $f_{ab}$ as the de Sitter metric. The reason behind this choice in contrast to the usual Minkowski metric is that one can easily be able to formulate the flat, open, or, closed cosmologies by suitable slicing of the de Sitter model. Also, this choice of reference metric eliminates the problem of `no-go' theorem \cite{DAmico1}. The metric in this model can be written as

\begin{equation}
f_{ab} d \phi^a d \phi^b= -dT^2+ b_K ^2 (T) \gamma_{ij} (X) dX^i dX^j,\label{eqn12}
\end{equation}

with

\begin{eqnarray}
b_0 (T)&=& e^{H_c T},\label{eqn13}\\
b_{-1} (T)&=& H_c ^{-1} \sinh(H_c T),\label{eqn14}\\
b_1(T)&=& H_c ^{-1} \cosh(H_c T).\label{eqn15}
\end{eqnarray}

It should be mentioned that in the limit, $H_c \longrightarrow 0$, one recovers Minkowski's metric for flat and open cases: $b_0= 1, b_{-1}= T$, while the later case corresponds to the Milne metric for flat geometry.

Further, specifying the Stueckelberg field as

\begin{equation}
\phi^0= T= f(t),~~~~~\phi^i= X^i= x^i,\label{eqn16}
\end{equation}

one sees that the cosmological symmetries are satisfied and from Eq. (\ref{eqn9}), $\Sigma_{\mu \nu}$ becomes a homogeneous and isotropic tensor of the form

\begin{equation}
\Sigma_{\mu \nu}= diag\left(-\dot{f}^2, b_K ^2 \left(f(t)\right) \gamma_{ij}\right). \label{eqn17}
\end{equation}

So, from Eq. (\ref{eqn7}), the elements of $\kappa$ matrix have the simple form as

\begin{eqnarray}
\kappa_0 ^0&=& 1- \zeta_f \frac{\dot{f}}{N},\label{eqn18}\\
\kappa_j ^i&=& (1-\frac{b_K (f)}{a}) \delta_j ^i,\label{eqn19}\\
\kappa^i _0&=& 0,\label{eqn20}\\
\kappa^0 _i&=& 0,\label{eqn21}
\end{eqnarray}

where $\zeta_f$ denotes the sign of `$\dot{f}$'. Thus, in the Lagrangian, the non-linear higher derivative term for massive gravity (denoting the potential for graviton) becomes

\begin{equation}
L_{mg}= \sqrt{-g} U (g, H)= \sqrt{-g} (U_2+ \alpha_3 U_3+ \alpha_4 U_4),\label{eqn21.1}
\end{equation}

or, equivalently,

\begin{eqnarray}
L_{mg}&=& (a- b_K (f)) \left[N \{a^2 (4 \alpha_3+ \alpha_4+ 6)-a (5 \alpha_3+ 2 \alpha_4+ 3) b_K (f)+ (\alpha_3+ \alpha_4) b^2 _K (f)\}\right.\nonumber\\
&&\left.- \zeta_f \dot{f} \{a^2 (3+ 3 \alpha_3+ \alpha_4)- a (3 \alpha_3+ 2 \alpha_4) b_K (f)+ \alpha_4 b^2_K (f)\}\right].\label{eqn22}
\end{eqnarray}

So, the variation of this Lagrangian with respect to `$f$' gives the equation of motion for $f (t)$ as

\begin{equation}
\left[(3+ 3 \alpha_3+ \alpha_4) a^2-2 (1+ 2 \alpha_3+ \alpha_4) a b_K (f)+ (\alpha_3+ \alpha_4) b^2 _K (f)\right] \left(\frac{\dot{a}}{N}- \zeta_f b_K ^\prime (f)\right)= 0. \label{eqn23}
\end{equation}

From the square bracket, we have

\begin{eqnarray}
b_K (f(t))&=& \mu_{\pm} a(t),\label{eqn24}\\
\mu_{\pm}&=& \frac{1+ 2\alpha_3+ \alpha_4 \pm \sqrt{1+ \alpha_3+ \alpha_3 ^2- \alpha_4}}{\alpha_3+ \alpha_4}.\label{eqn25}
\end{eqnarray}

So, one can obtain $f(t)$ from Eq. (\ref{eqn24}), provided, $b_k$ is invertible. It should be noted that, in case of Minkowski's reference metric (i.e., $f_{ab}= \eta_{ab}$), we have $b_0 (f)= 1$ for flat case, and hence, there is no longer any solution, while, for open case (where $b_{-1} (f)= f$), there are two branches of solutions. On the other hand, from the remaining part of Eq. (\ref{eqn23}), we have

\begin{equation}
\zeta_f b_K ^\prime (f)= \frac{\dot{a}}{N},\label{eqn26}
\end{equation}

and as before, non-trivial solutions are possible, only if `$b_K ^\prime$' is an invertible function. In this case, there is no analog of Minkowskian reference metric (as there exists no solution for both flat and open cases). However, with the choice of $b_K$ from Eqns. (\ref{eqn13}), (\ref{eqn14}), and (\ref{eqn15}), it is possible to have an explicit solution for $f (t)$. In particular, for flat case (i.e., $K= 0$) (also assuming, $\dot{f}> 0$) we have

\begin{eqnarray}
f (t)&=& H_c ^{-1} ln \left(\frac{H (t) a(t)}{H_c}\right),\label{eqn27}\\
H&=& \frac{1}{N} \frac{\dot{a}}{a},\label{eqn28}
\end{eqnarray}

where $H$ is the Hubble rate of the FLRW metric, and $H_c$ that of the reference metric (here, de Sitter metric).

\section{FLRW cosmology in massive gravity theory and the possible transitions of the deceleration parameter}

For the present FLRW metric, the usual Einstein--Hilbert (EH) term takes the form

\begin{equation}
L_{EH}= -\frac{3 \dot{a}^2 a}{N}+ 3 K N a,\label{eqn29}
\end{equation}

together with which we have an arbitrary matter Lagrangian $L_m$ that describes ordinary cosmological matter. Thus, the variation of the total Lagrangian with respect to the lapse function $N$ (which sets to unity in the following) gives first Friedmann's equation

\begin{equation}
3  \left(H^2+ \frac{K}{a^2}\right)= 8 \pi G (\rho_m+ \rho_g),\label{eqn30}
\end{equation}

where $\rho_m$ denotes the ordinary matter energy density, and, $\rho_g$ is considered as an effective energy density due to massive gravity action as follows

\begin{equation}
\rho_g= \frac{m_g ^2}{8 \pi G a^3} (b_K (f)- a) \left[(6+ 4 \alpha_3+ \alpha_4) a^2-(3+ 5 \alpha_3+ 2\alpha_4) a b_K (f)+ (\alpha_3+ \alpha_4) b_K^2 (f)\right].\label{eqn31}
\end{equation}

Further, the variation of the total Lagrangian with respect to $a (t)$ yields second Friedmann's equation in the form

\begin{equation}
2 \dot{H}+ 3 H^2+ \frac{K}{a^2}= -8 \pi G (p_m+ p_g),\label{eqn32}
\end{equation}

where $p_m$ is the thermodynamic pressure of the cosmic matter and the effective pressure $p_g$ takes the form

\begin{eqnarray}
p_g&=& \frac{m_g ^2}{8 \pi G a^2} \left[\{6+ 4 \alpha_3+ \alpha_4- (3+ 3 \alpha_3+ \alpha_4) \dot{f}\} a^2-2 \{3+ 3 \alpha_3+ \alpha_4- (1+ 2 \alpha_3+ \alpha_4) \dot{f}\} a b_K (f)\right.\nonumber\\
&&\left. +\{1+ 2 \alpha_3+ \alpha_4- (\alpha_3+ \alpha_4) \dot{f}\} b_K ^2 (f)\right].\label{eqn33}
\end{eqnarray}

As we are restricted to flat FLRW model, so, using the solution (\ref{eqn27}) for $f (t)$, the expressions for $\rho_g$ and $p_g$ are simplified to

\begin{eqnarray}
\rho_g&=& \frac{m_g ^2}{8 \pi G} \left(-6 \alpha+ 9 \beta \frac{H}{H_c}- 3 \gamma \frac{H^2}{H_c^2}+ 3 \delta \frac{H^3}{H_c^3}\right),\label{eqn34}\\
p_g&=& -\rho_g+ \frac{m_g ^2}{8 \pi G} \frac{\dot{H}}{H H_c} \left(-3 \beta+ 2 \gamma \frac{H}{H_c}- 3 \delta \frac{H^2}{H_c^2}\right).\label{eqn35}
\end{eqnarray}

Note that, the present massive gravity theory contains three free parameters $m_g$, $\alpha_3$, and $\alpha_4$; and the parameters $\alpha, \beta, \gamma$ and $\delta$ in Eqns. (\ref{eqn34}) and (\ref{eqn35}) are not independent, rather, they are related to $\alpha_3$ and $\alpha_4$ by the following relations:

\begin{eqnarray}
\alpha&=& 1+ 2\alpha_3+ 2 \alpha_4,\label{eqn36}\\
\beta&=& 1+ 3 \alpha_3+ 4 \alpha_4,\label{eqn37}\\
\gamma&=& 1+ 6 \alpha_3+ 12 \alpha_4,\label{eqn38}\\
\delta&=& \alpha_3+ 4 \alpha_4.\label{eqn39}
\end{eqnarray}

Interestingly, if $H(z)= H_c$, then there is no contribution of massive graviton to the energy density (i.e., $\rho_g= 0$). Using the density parameters, Friedmann equation (\ref{eqn30}) can be written as (with $K= 0$)\\

\begin{equation}
\Omega_m+ \Omega_g= 1,\label{eqn40}
\end{equation}

where $\Omega_g$ =$\rho_g/\rho_c$ ($\rho_c= 3 H^2/8 \pi G$) is termed as the density parameter corresponding to the induced matter field in the massive gravity theory. Also, from the above Friedmann Eqns. (\ref{eqn30}) and (\ref{eqn32}), the expression for the deceleration parameter looks \cite{Pan1}

\begin{equation}
q= \frac{\left[\frac{1}{2}\left(1+3 \Omega_m \omega_m\right)-\frac{m_g ^2}{2 H^2}\left(-6 \alpha+ 9 \beta \frac{H}{H_c}-3 \gamma \frac{H^2}{H_c ^2}+3 \delta \frac{H^3}{H_c ^3}\right)-\frac{1}{2 H H_c}\left(-3 \beta+ 2 \gamma \frac{H}{H_c}-3 \delta \frac{H^2}{H_c ^2}\right)\right]}{\left[1+\frac{1}{2 H H_c}\left(-3 \beta+ 2 \gamma \frac{H}{H_c}-3 \delta \frac{H^2}{H_c ^2}\right)\right]},\label{eqn41}
\end{equation}

where $\omega_m= p_m/\rho_m$, is the equation of state parameter for the given fluid.\\

FIG. 1 shows the graphical representation of $q$ over the Hubble parameter ($H$) for the mentioned set of parameters in the graph. We also mention that, for two very very small values of the graviton mass $m_g= 0.0000001$ (dashed line), and $m_g= 0.00001$ (dotted line), the graphs overlap with one another, and even for $m_g= 0.01$, we can not distinguish between two graphs. It indicates the complete evolution of our universe from inflation to late time acceleration, where we have used the recent value of the density parameter $\Omega_m$ \cite{ade1}. The right and left tails of the curves indicate the inflationary and the late-time accelerating phase of the universe, while, the matter dominated era corresponds to the middle portion of the graph, where $q> 0$. Moreover, from the graph, we see that, if the graviton mass ($m_g$) is very very small, then the model approaches $\Lambda$CDM, however, if $m_g$ increases, then the null energy condition is violated at late-times, and this may be potentially problematic for the stability of the theory. As the reference metric is de Sitter, and $H_c$ corresponds to its Hubble rate, so, $H_c$ would define where inflation may occur. However, in the present graphs, the chosen value of $H_c (= 1)$ falls into the matter dominated era.

\begin{figure}
\begin{minipage}{0.5\textwidth}
\includegraphics[width= 1.1\linewidth]{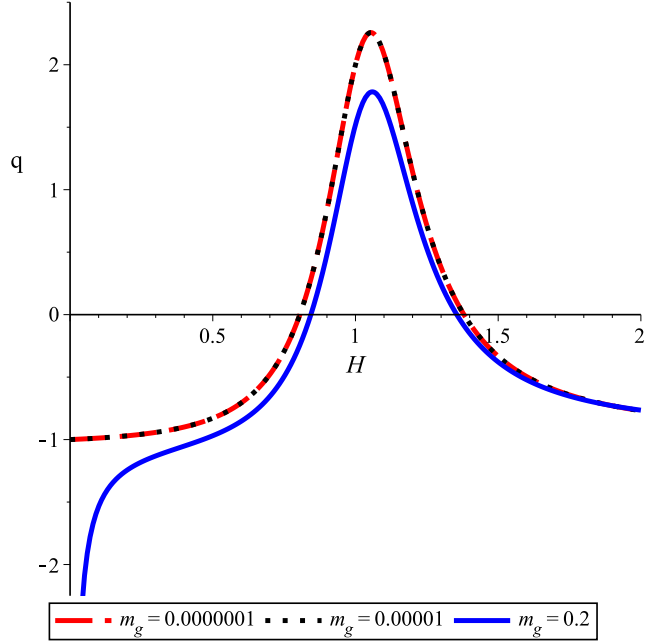}
\caption{The figure shows the cosmic evolution from inflation to late time acceleration (see Eq. (\ref{eqn41})) for the following parameterizations: $\alpha_3= -1$; $\alpha_4= -2$, $\omega_m= 0$, $\Omega_m= 0.3086$ and $H_c= 1$.}
\end{minipage}
\end{figure}

\section{Equivalence of Massive gravity theory and Einstein gravity: Cosmological solutions}

In the previous section, we have presented effective Friedmann equations in massive gravity theory containing three free parameters. Here, we start with restricting the free parameters $\alpha_3$, $\alpha_4$, such that, $\beta= \delta= 0$. As a result, we have $\alpha_3= - \frac{1}{2}$, $\alpha_4= \frac{1}{8}$, thus, we have $\alpha= \frac{1}{4}$, and $\gamma= -\frac{1}{2}$.

So, the Friedmann Eqns. (\ref{eqn30}) and (\ref{eqn32}) are now simplified to

\begin{eqnarray}
3 H^2&=& 8 \pi G \rho_m- \frac{3}{2} m_g ^2+ \frac{3 m_g ^2}{2 H_c ^2} H^2,\label{eqn42}\\
2 \dot{H}&=& -8 \pi G (\rho_m+ p_m)+ \frac{m_g ^2}{H_c ^2} \dot{H},\label{eqn43}
\end{eqnarray}

which can again be rewritten as

\begin{eqnarray}
3 H^2&=& 8 \pi G_m \rho_m -\Lambda,\label{eqn44}\\
2 \dot{H}&=& -8 \pi G_m (\rho_m+ p_m),\label{eqn45}
\end{eqnarray}

with $G_m$ and $\Lambda$ as

\begin{eqnarray}
G_m&=& \frac{G}{\left(1-\frac{m_g ^2}{2 H_c ^2}\right)},\label{eqn46}\\
\Lambda&=& \frac{3 m_g ^2}{2 \left(1-\frac{m_g ^2}{2 H_c ^2}\right)}.\label{eqn47}
\end{eqnarray}

Note that, Eqns. (\ref{eqn44}) and (\ref{eqn45}) are nothing but usual Friedmann's equations in Einstein gravity with a negative cosmological constant which depends on the graviton mass, and also, the Newtonian gravitational constant is modified by the free parameter $m_g$. It should be mentioned that, if the graviton is assumed to be massless, then we get back usual Newton's gravitational constant, and the cosmological constant vanishes. In the literature \cite{shapiro1, jw1, ghs1, jw2, maeda1}, it is claimed that negative cosmological constant has a significant role in describing the evolution of the universe. The low energy limit of supersymmetry prefers a negative cosmological constant implying an Anti de Sitter (AdS) cosmos \cite{van1}. Also, the problem with negative cosmological constant in supersymmetry theory can be resolved using Einstein--Cartan theory \cite{boehmer1}. It was speculated by Biswas et al. \cite{biswas1}, that our universe was begun with a negative cosmological constant. However, Kallosh and Linde \cite{kallosh1} claimed that in presence of a negative cosmological constant, our universe may recollapse in future.

Now, we shall present some cosmological solutions in massive gravity theory considering its equivalence to Einstein gravity with negative cosmological constant.

\subsection{When matter in the form of a perfect fluid with constant equation of state}

If the matter is chosen as a perfect fluid with constant equation of state (EoS) as, $p_m= \omega_m \rho_m$; $\omega_m$ is a constant, then the Hubble parameter and the scale factor can be obtained by solving the field equations (\ref{eqn44}) and (\ref{eqn45}) as

\begin{eqnarray}
H&=& \sqrt{\frac{|\Lambda|}{3}} tan \left[\sqrt{\frac{|\Lambda|}{3}} \left(A- \epsilon_0 t\right)\right],\label{eqn48}\\
a&=& B \left|cos \left( \sqrt{\frac{|\Lambda|}{3}}\left(A- \epsilon_0 t\right)\right)\right|^{\frac{1}{\epsilon_0}}.\label{eqn49}
\end{eqnarray}

Also, the deceleration parameter ($q$) takes the form

\begin{equation}
q= -1+ \epsilon_0 \left[1+ cot^2 \left(\sqrt{\frac{|\Lambda|}{3}} \left(A- \epsilon_0 t\right)\right)\right].\label{eqn50}
\end{equation}

Here, $A$, $B$ are the constants of integration, and, $\epsilon_0$ is the slow roll parameter in the case when $\Lambda= 0$, i.e., $\epsilon_0$:= $- \left(\frac{\dot{H}}{H^2}\right)_{\Lambda \rightarrow 0}= \frac{3}{2} (1+ \omega_m)$. The above solutions represent a model of the cyclic universe compared to the power law expanding solution in standard cosmology provided $\left(\frac{8 \pi G_m}{3}\right) \rho_0- \frac{\Lambda}{3}> 0$, i.e., if the initial energy density of the fluid ($\equiv \rho_0$) is large enough to compensate the negative cosmological constant \cite {prokopec1}, and $\rho_0$ depends on the graviton mass.

Now, the expression of the Ricci scalar

\begin{equation}
R \equiv 6\left(\dot{H}+ 2 H^2\right)= 2 |\Lambda| \left[\left(2- \epsilon_0\right) tan^2 \left(\sqrt{\frac{|\Lambda|}{3}}(A- \epsilon_0 t)\right)- \epsilon_0\right],\label{eqn51}
\end{equation}

shows that $R$ diverges at $t_d= \frac{1}{\epsilon_0} \left[A+ \sqrt{\frac{3}{|\Lambda|}} (2n +1) \frac{\pi}{2}\right]$, $n \in {\bf Z}$ (set of integers), and it is the key feature of the cyclic universe, i.e., a universe with successive expansion and contraction, and continues in this manner, and this `$t_d$' depends on the massive gravity effect through the cosmological constant in Eq. (\ref{eqn47}). Thus, this cyclic nature of the universe comes from the massive gravity effect.

\subsection{When the fluid obeys the Equation of state (EoS) for Chaplygin gas}

The EoS for the Chaplygin gas is $p= -\frac{A}{\rho}$, where $A> 0$. Now, inserting this EoS in Eq. (\ref{eqn45}), and integrating we have

\begin{equation}
H^2= -8 \pi G_m \sqrt{A} \int \left[\left(1+ \frac{D^2}{A a^6}\right)^{-\frac{1}{2}}- \left(1+ \frac{D^2}{A a^6}\right)^{\frac{1}{2}}\right] \frac{da}{a}.\label{eqn52}
\end{equation}

Here, we consider the solution for large $a$. Therefore, using binomial expansion, Eq. (\ref{eqn52}) gives a solution for the Hubble parameter as

\begin{equation}
H= \left(\sqrt{\frac{4 \pi D^2 G_m}{3 A^{\frac{1}{2}}}} \right) \frac{1}{a^3},\label{eqn53}
\end{equation}

and, consequently, solving Eq. (\ref{eqn53}), we have the scale factor as

\begin{equation}
a^3= \left(\sqrt{\frac{12 \pi D^2 G_m}{A^{\frac{1}{2}}}} \right) t+ E.\label{eqn54}
\end{equation}

The deceleration parameter $q$ takes the form

\begin{equation}
q\equiv -\left(1+ \frac{a}{H^2} \frac{dH}{da}\right)= -1+ \left(3 \sqrt{\frac{6 A^{\frac{1}{2}}}{8 \pi D^2 G_m}} \right) \frac{1}{a^3}.\label{eqn55}
\end{equation}

Here $D$ and $E$ in the above equations are the constants of integration. Hence, we have a power law expansion of the universe, and the model gradually approaches $\Lambda$CDM ($q= -1$) for large $a$.\\

Further, for this perfect fluid model, the general field Eqns. (\ref{eqn30}) and (\ref{eqn32}) can be rewritten as (for simplicity, we assume,  $8 \pi G= c= 1$.)

\begin{eqnarray}
3 H^2 (1+ \gamma \frac{m_g ^2}{H_c ^2})&=& \rho_0 a^{-3(1+\omega_m)}+ 3 m_g ^2 \left(-2 \alpha+ 3 \beta \frac{H}{H_c}+ \delta \frac{H^3}{H_c ^3}\right),\label{eqn56}\\
\dot{H}&=& -\frac{\rho_0 H H_c (1+\omega_m)a^{-3(1+\omega_m)}}{\left[2 H H_c+ m_g ^2 \left(-3 \beta+2 \gamma \frac{H}{H_c}-3 \delta \frac{H^2}{H_c ^2}\right)\right]},\label{eqn57}
\end{eqnarray}

where $\rho_m$ is eliminated by solving the matter conservation equation. Thus, evolution Eq. (\ref{eqn57}) with the definition of the Hubble parameter

\begin{equation}
\dot{a}= a H,\label{eqn58}
\end{equation}

together form an autonomous system in the phase plane ($a$, $H$), provided, $1+\omega_m \neq 0$, i.e., the fluid model is not a $\Lambda$CDM model. Note that $a$ and $H$ are not constants rather they are constrained by the relation (\ref{eqn56}). The critical points of the dynamical system are given for the following two regions\\

$\star$ when the fluid is not in the phantom region, i.e., $1+ \omega_m> 0$. The critical points are

\begin{equation}
(H= 0, a= a_0);~~~~~~~~~~~a_0= \left(\frac{\rho_0}{6 \alpha m_g ^2}\right)^{\frac{1}{3(1+\omega_m)}},\label{eqn59}
\end{equation}

$\star$ when there is phantom fluids, i.e., $1+\omega_m= 0,$

\begin{equation}
(H= H_0, a= 0).\label{eqn60}
\end{equation}

The parameter $H_0$ in Eq. (\ref{eqn60}) satisfies the cubic equation

\begin{equation}
H^3- \frac{(H_c ^2+ \gamma m_g ^2)}{m_g ^2 \delta} H^2+ \frac{3 \beta H_c ^2}{\delta} H- \frac{2 \alpha H_c ^3}{\delta}= 0.\label{eqn61}
\end{equation}

As $H_0$ is the real root of the cubic Eq. (\ref{eqn61}), so the number of critical points depends on the choice of the parameters $\alpha_3, \alpha_4, m_g$, and $H_c$.

\section{Emergent Scenario}

The proposed cosmological scenarios to avoid the initial singularity (big bang) of the standard cosmology can be classified as bouncing universes, or, the emergent universes. In this section, we shall examine whether in the context of massive gravity theory, it is possible, or not, to have the emergent scenario which arises due to the search for the singularity free inflationary models in the framework of standard cosmology. Briefly, emergent universe is a model universe having no time-like singularity, ever existing and almost static behavior in the infinite past (t$\longrightarrow$ $-\infty$). It is worthwhile to mention that, in 1988, Brandenberger and Vafa \cite{BV1988} showed that in the context of superstring theory, a simple model can resolve this big bang singularity. After that, Ellis and Maartens \cite{EM2004}, and Ellis et al. \cite{Ellis2004} discussed the same scenario of the universe to remove this big bang singularity. Subsequently, Mukherjee et al. \cite{Mukherjee2005} found some solutions for Starobinsky model having features of an emergent universe. Also, Mukherjee et al. \cite{Mukherjee2006} formulated a general framework for an emergent universe considering ``GR+ exotic matter''. As a result, several works were done on the emergent universe \cite{Mulryne2005, Nunes2005, Lidsey2006, Banerjee2007, Banerjee2008}. Recently, in the context of non-equilibrium thermodynamics, gravitationally induced particle creations exhibit the emergent universe scenario \cite{Chakraborty2014}.

In the context of massive gravity theory, we start with the Friedmann Eqns. (\ref{eqn44}) and (\ref{eqn45}). These equations can be interpreted as a single fluid in Einstein gravity as

\begin{eqnarray}
3 H^2&=& 8 \pi G \rho_{eff},\label{eqn62}\\
2 \dot{H}&=& -8 \pi G (\rho_{eff}+p_{eff}),\label{eqn63}
\end{eqnarray}

where

\begin{eqnarray}
\rho_{eff}&=& \frac{\rho_m- \frac{\Lambda}{8 \pi G}}{(1-\frac{m_g ^2}{2 H_c ^2})},\label{eqn64}\\
p_{eff}&=& \frac{p_m+ \frac{\Lambda}{8 \pi G} }{(1-\frac{m_g ^2}{2 H_c ^2})}.\label{eqn65}
\end{eqnarray}

The energy conservation equation

\begin{equation}
\dot{\rho}_m+ 3H (\rho_m+p_m)= 0,\label{eqn66}
\end{equation}

now becomes

\begin{equation}
\dot{\rho}_{eff}+ 3H (\rho_{eff}+ p_{eff})= 0.\label{eqn67}
\end{equation}

If we now assume the equation of state parameter of the effective fluid as

\begin{equation}
p_{eff}= -\frac{\epsilon}{H} \rho_{eff},\label{eqn68}
\end{equation}

where $\epsilon > 0$ is some constant, but depends on the mass of the graviton. Now, integrating (\ref{eqn67}) with the assumption in (\ref{eqn68}) we have

\begin{equation}
\rho_{eff}= \frac{\rho_0}{a^3} e^{3 \epsilon t}. \label{eqn69}
\end{equation}

Further, using (\ref{eqn68}) in the Friedmann Eqns. (\ref{eqn62}) and (\ref{eqn63}), we have

\begin{equation}
2 \dot{H}= -3 H^2+3 \epsilon H,\label{eqn70}
\end{equation}

which on integration gives

\begin{equation}
H= \frac{3 \epsilon}{3+ e^{-\tau}},\label{eqn71}
\end{equation}

where $\tau= \frac{3 \epsilon}{2} (t-t_0)$, and $t_0$ is the constant of integration. Now, integrating the above relation once more, one gets the scale factor for the emergent scenario as

\begin{equation}
\left(\frac{a}{a_0}\right)^{\frac{3}{2}}= 1+ 3 e^\tau ,~~~\mbox{with}~~a_0= a(t_0).\label{eqn72}
\end{equation}

From the above solution in Eq. (\ref{eqn72}), we see that, as $t \longrightarrow -\infty$, $a \longrightarrow a_0$, and, for $t \ll t_0$, $a \simeq a_0$, while the scale factor a grows exponentially for $t> t_0$. In particular, the above cosmological solution has the following asymptotic behavior:\\

(I) a $\longrightarrow$ $a_0$, H$\longrightarrow$ 0, as t$\longrightarrow$ $-\infty$\\

(II) a $\simeq$ $a_0$, H $\simeq$ 0, for t$<< t_0$\\

(III) a $\simeq$ $e^{H_0 (t-t_0)}$, for t$>> t_0$.\\

Hence, it is possible to have an emergent scenario in massive gravity theory for a perfect fluid with variable equation of state (given in Eq. (\ref{eqn68})) depending on the mass of the graviton.

\section{Cosmography of Massive Gravity}

In cosmography analysis, the universe is assumed to be homogeneous and isotropic on the largest scale and no specific dynamical theory is assumed a priori. The scale factor is expanded in Taylor series with respect to the cosmic time as

\begin{eqnarray}
\frac{a(t)}{a(t_0)}= 1+ H_0 (t- t_0)- \frac{1}{2!} q_0 H_0 ^2 (t- t_0)^2+  \frac{1}{3!} j_0 H_0 ^3 (t- t_0)^3 +\frac{1}{4!} s_0 H_0 ^4 (t- t_0)^4 &+&
\nonumber
\\
\frac{1}{5!} l_0 H_0 ^ 5 (t- t_0)^5 +\frac{1}{6!} m_0 H_0 ^6 (t- t_0)^6+ O(|t- t_0|^7); \label{eqn73}
\end{eqnarray}

where $t_0$ is the present time, and, the suffix `$0$' indicates the value of the quantity at present. The above coefficients of different powers of `$t$' are defined as \cite{visser1}

$$H= \frac{\dot{a}}{a},~q= -\frac{1}{a H^2} \frac{d^2 a}{dt^2},~j= \frac{1}{aH^3} \frac{d^3a}{dt^3},~s= \frac{1}{aH^4} \frac{d^4a}{dt^4},~l= \frac{1}{aH^5} \frac{d^5a}{dt^5},~m=\frac{1}{aH^6} \frac{d^6a}{dt^6};$$

and are conventionally termed as Hubble, deceleration, jerk ($j$), snap ($s$), lerk ($l$) and $m$ parameters respectively. These parameters can be used to find the distance-redshift relation and hence different distances in the universe. Further, the sign of $q$ indicates acceleration ($-$ve), or, deceleration ($+$ ve), similarly, a change of the sign of $j$ (in an expanding universe) signals that the acceleration starts increasing or decreasing.

In this context, we have studied the cosmographic parameters for three different values of the graviton mass ($m_g$).

FIGs. 2--5 show the variations of the four cosmographic parameters $j, s, l$ and $m$ against the Hubble parameter ($H$). In FIG. 2, as there are several transitions of $j$, so the acceleration is fluctuating throughout the evolution as reflected from FIG. 1.

\begin{figure}
\begin{minipage}{0.4\textwidth}
\includegraphics[width= 1.1\linewidth]{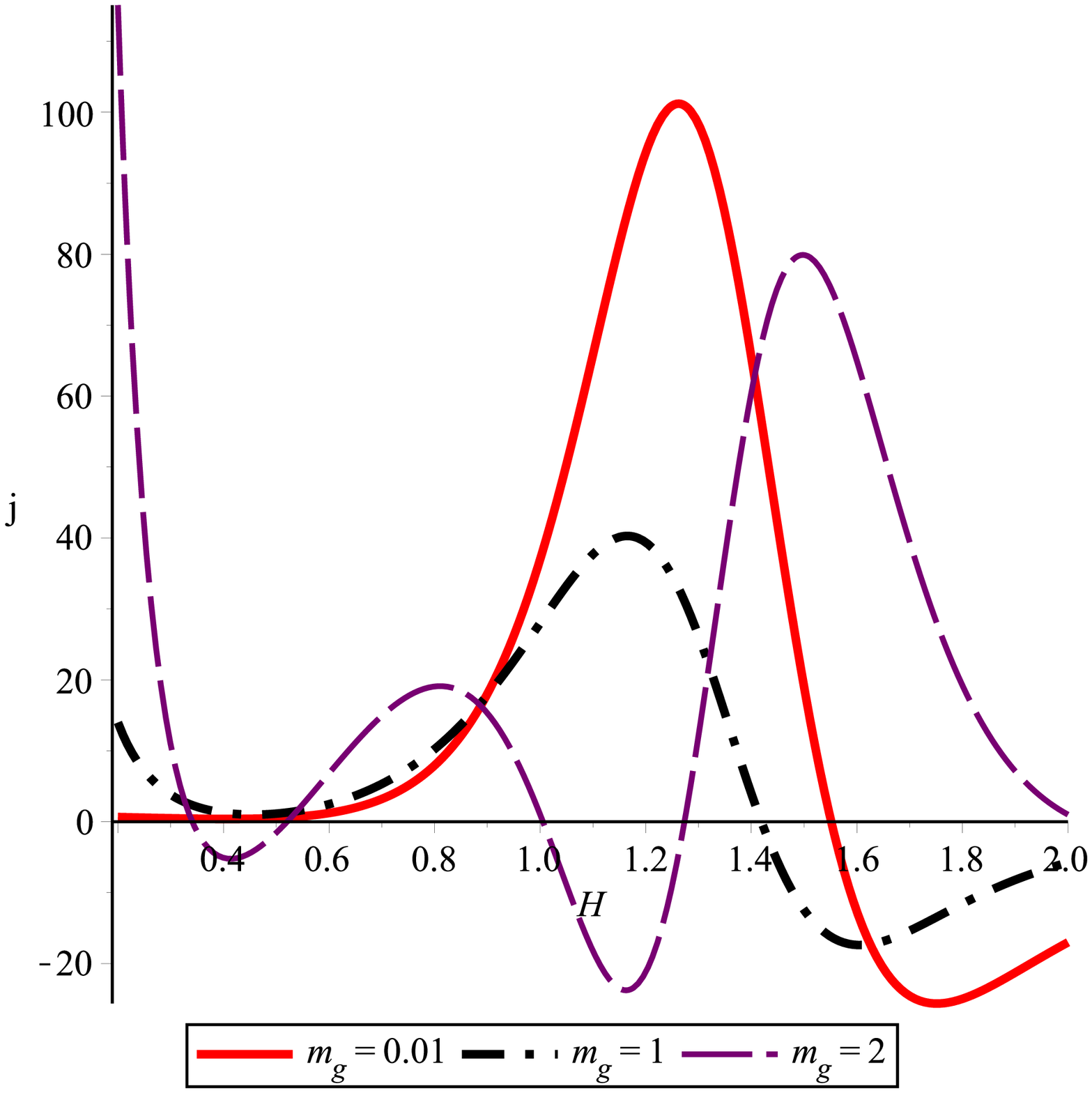}
\caption{This is a study of the cosmographic parameter $j$ described in the paper for three different graviton mass $m_g$. Parameters are $\alpha_3= -1$, $\alpha_4= 0$, $H_c= 1$ and $\omega_m= 0$ and $\Omega_m= 0.3086$ \cite{ade1}.}
\end{minipage}
\begin{minipage}{0.45\textwidth}
\includegraphics[width= 1.0\linewidth]{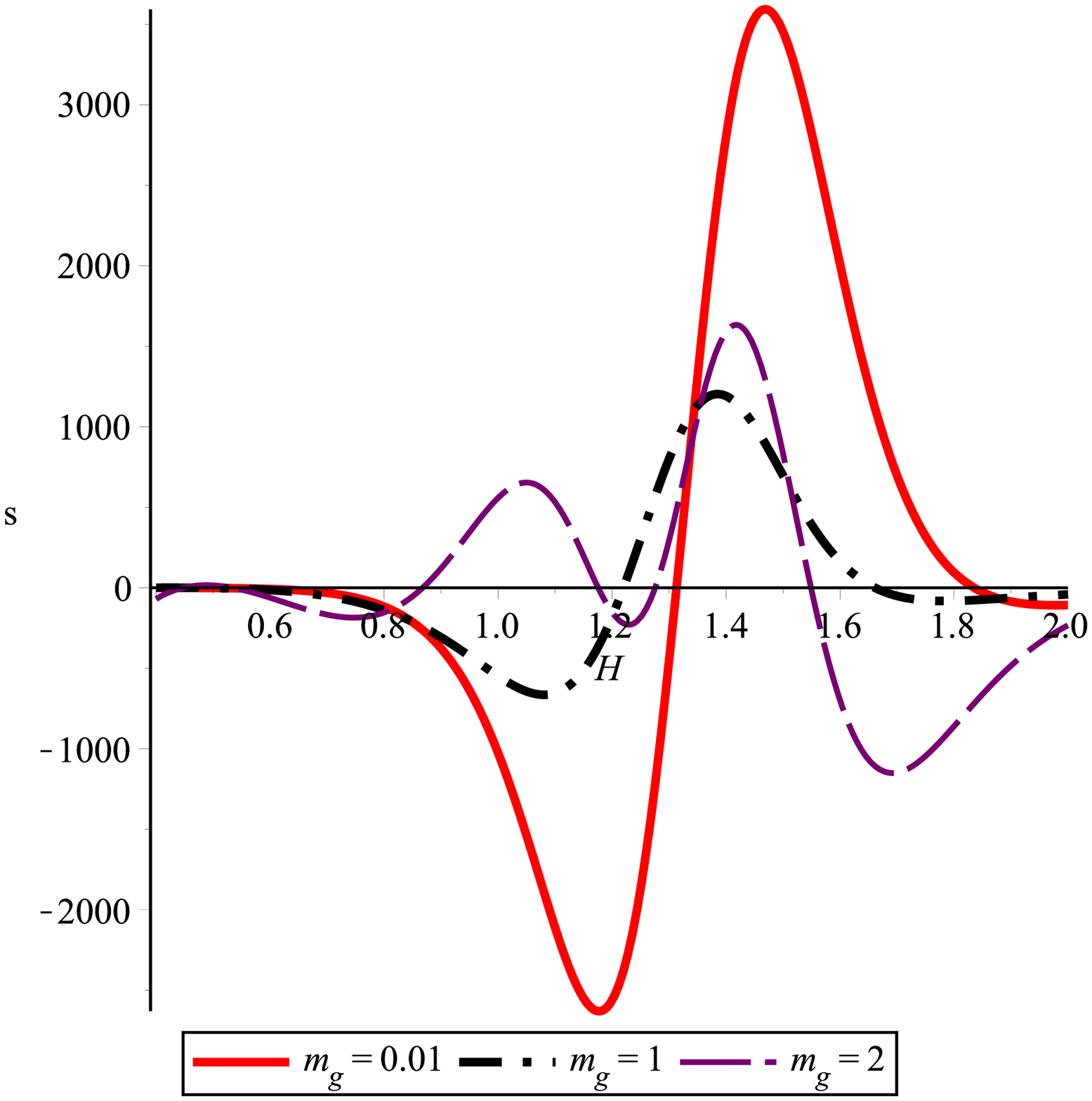}
\caption{A study of the cosmographic parameter $s$ described in the paper for three different graviton mass $m_g$. Parameters are $\alpha_3= -1$, $\alpha_4= 0$, $H_c= 1$ and $\omega_m= 0$ and $\Omega_m= 0.3086$ \cite{ade1}.}
\end{minipage}
\end{figure}

\begin{figure}
\begin{minipage}{0.4\textwidth}
\includegraphics[width= 1.1\linewidth]{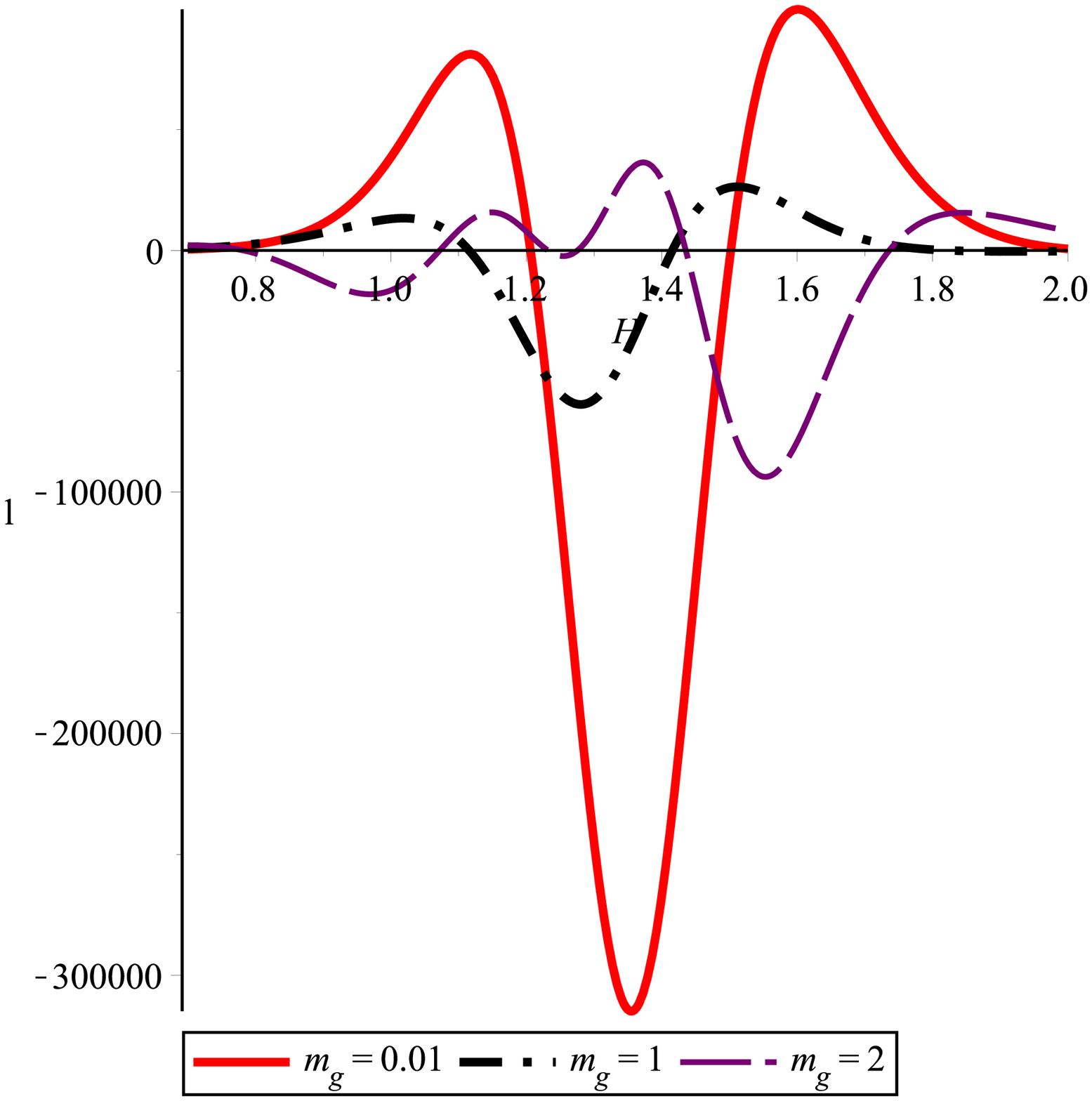}
\caption{This is studied for the cosmographic parameter $l$ described in the paper for three different graviton mass $m_g$. Parameters are $\alpha_3= -1$, $\alpha_4= 0$, $H_c= 1$ and $\omega_m= 0$ and $\Omega_m= 0.3086$ \cite{ade1}.}
\end{minipage}
\begin{minipage}{0.45\textwidth}
\includegraphics[width= 1.0\linewidth]{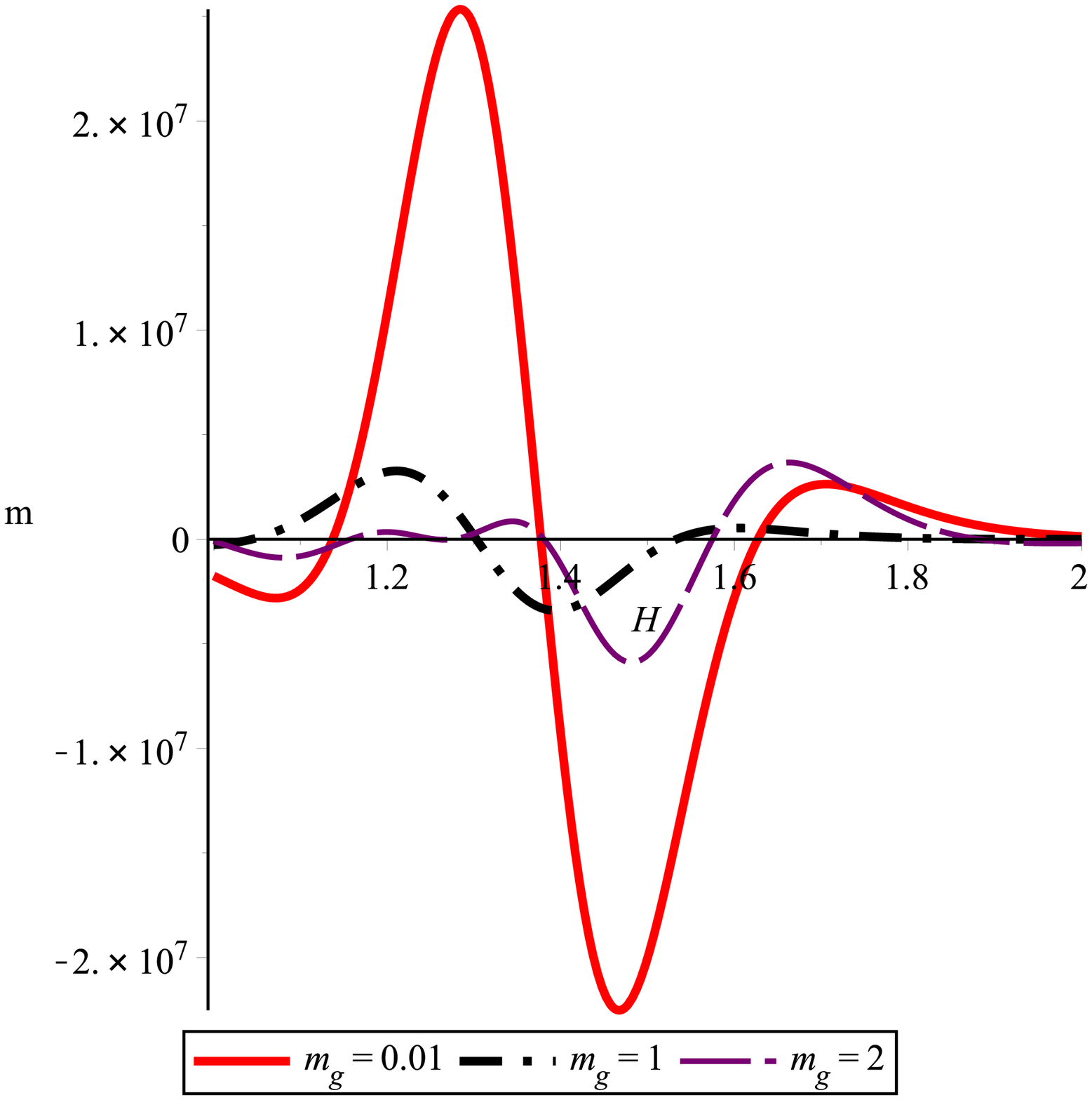}
\caption{It is a study of the cosmographic parameter $m$ described in the paper for three different graviton mass $m_g$. Parameters are $\alpha_3= -1$, $\alpha_4= 0$, $H_c= 1$ and $\omega_m= 0$ and $\Omega_m= 0.3086$ \cite{ade1}.}
\end{minipage}
\end{figure}

\section{Conclusions and outlook}

In massive gravity theory, we have studied several cosmological aspects. The background metric is taken as flat FLRW metric, and the matter is chosen as perfect fluid with constant equation of state. The deceleration parameter has been calculated, and FIG. 1 shows the transition from deceleration to present late time acceleration as predicted by the recent observations. Moreover, from FIG. 1, it has been found that the model will remain stable as long as the graviton mass is very small. However, the model gradually becomes unstable as the mass of the graviton gradually increases. Among the three free parameters in the massive gravity theory, by choosing $\alpha_3$ and $\alpha_4$ appropriately, it has been shown that the Friedmann's equations in this theory are equivalent to those in Einstein gravity with a negative cosmological constant, and, the Newton's gravitational constant is modified by the remaining free parameter (the mass of the graviton $m_g$). Thus, the choice of the free parameters $\alpha_3$ and $\alpha_4$ changes the qualitative feature of the massive gravity theory, and it is in difference with the work by Fasiello and Tolley \cite{FT2012}. The interpretation of negative cosmological constant has been discussed. A possibility of emergent scenario has been investigated in massive gravity theory. It is found that for a specific variable equation of state (related to the mass of the graviton) with negative pressure, it is possible to have this non-singular aspect at early epoch which was also obtained in the superstring theory \cite{BV1988}. Lastly, the cosmographic analysis has been done for this present gravity theory.

The cosmological solution for perfect fluid with constant equation of state shows a periodic nature of the scale factor which may be interpreted as the possibility of a cyclic universe in massive gravity theory. This feature is also supported by the divergence of the Ricci scalar at infinite number of points. Also, it should be mentioned that, this cyclic nature of the universe which comes from the massive gravity effect can be realized in the superstring theory as discussed by Brandenberger and Vafa in Ref. \cite{BV1988}. For Chaplygin gas model, the solution shows a power law expansion of the universe, and the model as expected approaches the $\Lambda$CDM model. Finally, the negative cosmological constant can be incorporated into the matter sector by introducing effective density and pressure in the Einstein equations.

\section{Acknowledgments}
SP acknowledges CSIR, Govt. of India for financial support through SRF scheme (File No. 09/096 (0749)/2012-EMR-I). SC thanks UGC-DRS programme at the Department of Mathematics, Jadavpur University. The authors are grateful to the anonymous referee, whose comments helped them to improve the manuscript considerably.

\end{document}